\documentclass[prb, twocolumn,showpacs,superscriptaddress]{revtex4}
\usepackage{amsfonts}
\usepackage{epsfig}
\usepackage{bm}
\usepackage{amsmath}
\usepackage{graphicx}
\usepackage{float}
\usepackage{hyperref}
\usepackage[english]{babel}

\begin{document}

\title{Critical point for the CAF-F phase transition at charge neutrality in bilayer graphene}
\author{S.Pezzini}

\affiliation{Dipartimento di Fisica, Universit\`{a} degli studi di Pavia, I-27100 Pavia, Italy}
\affiliation{Laboratorio de Bajas Temperaturas, Universidad de Salamanca, E-37008 Salamanca, Spain}

\author{C. Cobaleda}

\affiliation{Laboratorio de Bajas Temperaturas, Universidad de Salamanca, E-37008 Salamanca, Spain}
\affiliation{NEST, Scuola Normale Superiore, Piazza S. Silvestro 12, I-56127 Pisa, Italy}

\author{B. A. Piot}

\affiliation{Laboratoire National des Champs Magn\'{e}tiques Intenses, CNRS-UJF-UPS-INSA, F-38042 Grenoble, France}

\author{V. Bellani}

\affiliation{Dipartimento di Fisica, Universit\`{a} degli studi di Pavia, I-27100 Pavia, Italy}

\author{E. Diez}

\affiliation{Laboratorio de Bajas Temperaturas, Universidad de Salamanca, E-37008 Salamanca, Spain}

\begin{abstract}
We report on magneto-transport measurements up to 30 T performed on a bilayer graphene Hall bar, enclosed by two thin hexagonal boron nitride flakes. Our high mobility sample exhibits an insulating state at neutrality point which evolves into a metallic phase when a strong in-plane field is applied, as expected for a transition from a canted antiferromagnetic to a ferromagnetic spin ordered phase. For the first time we individuate a temperature-independent crossing in the four-terminal resistance as a function of the total magnetic field, corresponding to the critical point of the transition. We show that the critical field scales linearly with the perpendicular component of the field, as expected from the underlying competition between the Zeeman energy and interaction-induced anisotropies. A clear scaling of the resistance is also found and an universal behavior is proposed in the vicinity of the transition.
\end{abstract}


\maketitle

Bilayer graphene presents a unique energy spectrum upon the application of a perpendicular magnetic field \cite{McCann, Nov}, with different possible competing orders resulting from the large number of underlying symmetries of the system \cite{kharprl}. A huge experimental effort was dedicated to the identification of the anomalous insulating ground state at the charge neutrality point (CNP) in high-quality samples \cite{jairo, yacoby,freitag, weitz}. Recent theoretical \cite{kharprl, kharprb} and experimental results \cite{maher} support the formation of a canted antiferromagnetic (CAF) phase in perpendicular-only magnetic fields, which results from the competition between the Zeeman energy $\epsilon_Z=g\mu_B B_{tot}$ and an anisotropy energy $u_\perp$ originating from electron-electron or electron-phonon interaction at the lattice scale. Within the CAF ordering, the spin polarizations associated to the two valleys (and equivalently to the two sublattices and layers) have equal component in the direction of the magnetic field and are opposed in the perpendicular plane, with an optimal canting angle determined by the ratio $\epsilon_Z/u_\perp$. The energy spectrum of the CAF phase presents a gap both for the bulk and edge excitations, leading to the insulating transport behavior. While $\epsilon_Z$ depends on the total magnetic field, $u_\perp$ is sensitive only to the perpendicular component. The application of a strong in-plane field thus favors $\epsilon_Z$ and results in a smooth transition to a ferromagnetic (F) spin-ordered phase, which is distinctively metallic owing to the presence of gapless symmetry-protected edge states, with properties analogous to the ones of the quantum spin Hall (QSH) effect in two-dimensional topological insulators \cite{kanemele, hasan}.  

The critical point for the CAF-F transition is expected to be realized when the condition 
\begin{equation}
\label{trans}
\epsilon_Z(B_{tot})=2|u_\perp(B_\perp)|
\end{equation}
is satisfied at some specific $B_{\perp}$-dependent value of the total field $B^*_{tot}(B_{\perp})$, at which the edge energy gap vanishes \cite{kharprb}. While compelling evidence of the CAF-F transition has been reported in Ref.\onlinecite{maher} (and analogously in Ref.\onlinecite{afy} also for single-layer graphene), the experimental observation of its critical point has not been reported so far. Here we individuate and analyze the critical point for the the CAF-F transition at CNP in bilayer graphene. The data of the four-terminal resistance at CNP ($R^{CNP}_{xx}$) measured as a function of $B_{tot}$ show a clear T-independent crossing point corresponding to the critical field $B^*_{tot}$. By repeating the measurements at different fixed values of $B_{\perp}$, we show that $B^*_{tot}$ scales linearly with $B_{\perp}$, which gives further evidence to the scenario of spin-ordering at CNP governed by the $\epsilon_Z$-$u_\perp$ competition and allows to determine the energy-scale of $u_\perp$ itself. The observation of the critical point also allowed us to conduct a successful scaling analysis on the data and determine a universal expression for $R^{CNP}_{xx}$. 

\begin{figure}[!t]
\centerline{\includegraphics[width = 9cm]{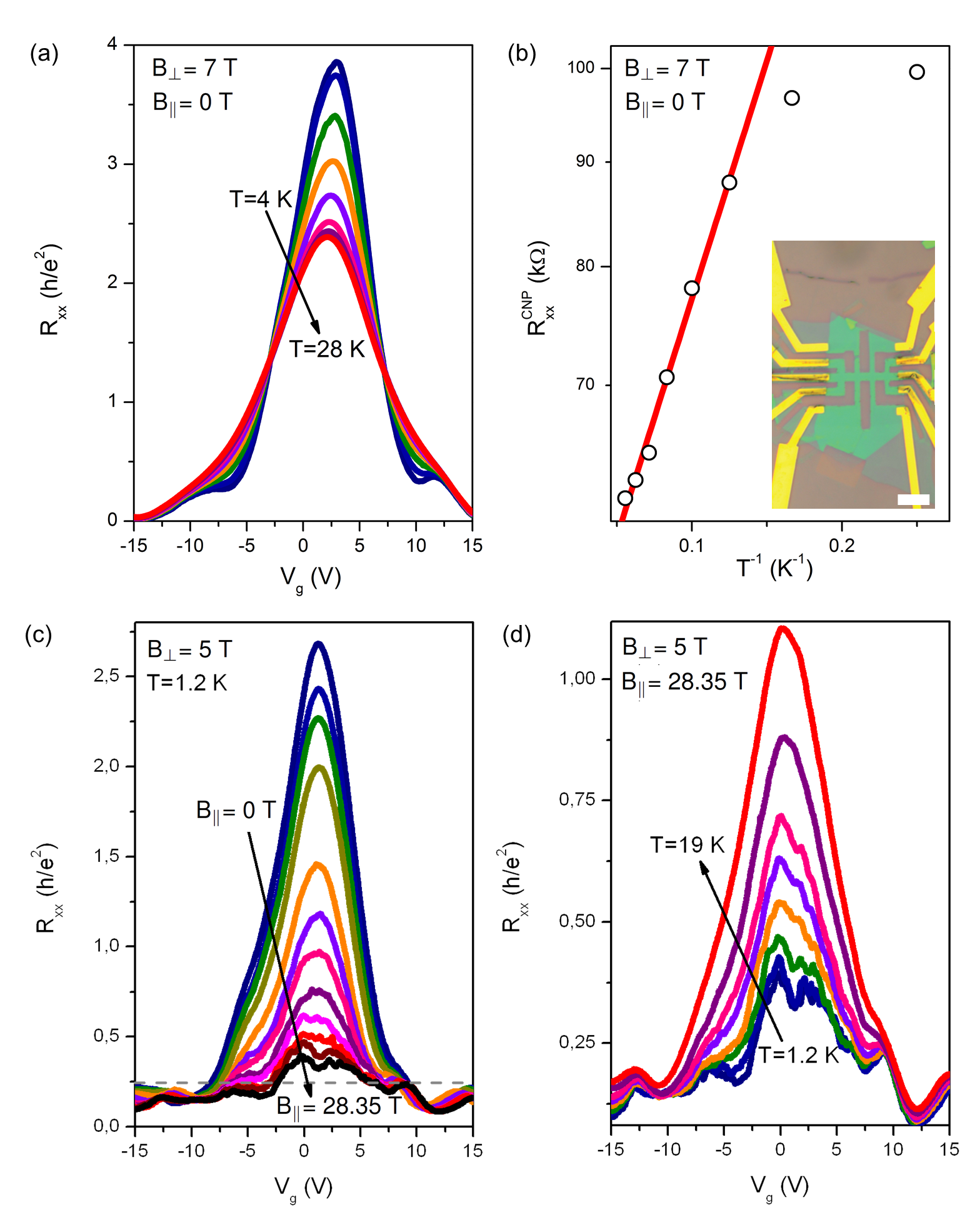}}
\caption{(Color online) (a) $R_{xx}$ as a function of gate voltage ($V_g$) for increasing temperatures at $B_\perp=B_{tot}=7$ T. (b) Temperature dependence of $R^{CNP}_{xx}$ in the CAF phase: the continuous red line is a fit to activated behaviour $R_{xx}^{CNP}\propto \exp{(\Delta/2k_BT)}$. (inset) Optical microscopy image of the sample; the scale bar corresponds to 5 $\mu$m. (c) $R_{xx}$ as a function of $V_g$ for increasing values of in-plane field with fixed perpendicular component ($B_{\perp}=5$ T), measured at $T\simeq 1.2$ K. The dashed line indicates the theoretical values $R_{xx}=\frac{1}{4}h/e^2$ expected for the four QSH-like states associated to the F phase at CNP in bilayer graphene. (d) $R_{xx}$ as a function of $V_g$ for increasing temperature in the F phase}
\label{one}
\end{figure}

The sample studied (see inset of \figurename~\ref{one} (b)) is a van der Waals heterostructure consisting in a stacking of hexagonal boron nitride ($h$-BN)\cite{dean}, bilayer graphene and $h$-BN, produced using the pick-up technique described in Ref.\onlinecite{niko}; an analogous sample (with an additional top gate contact) was studied in Ref.\onlinecite{maher}. Ti/Au contacts were evaporated on bilayer graphene, while $h$-BN, acting as a high-quality dielectric material, allows high electron mobility ($\mu\sim 5\times 10^4$ $cm^2V^{-1} s^{-1}$)\cite{cob}. We performed electrical transport measurements with standard low-frequency ($\approx 13$ Hz) ac lock-in technique, using an excitation current of 100 nA and varying the carrier density with a back-gate contact. The sample was mounted on a holder with rotation capability, allowing to vary the angle between the magnetic field and the graphene plane continuously from $0^\circ$ to $180^\circ$. The correct orientation of the sample was tested after each modification of the angle by carefully checking the position of the integer quantum Hall plateaus at filling factor $\pm 4$, which depends on the $B_{\perp}$ component only. In addition, the sample was oriented so that the current was flowing parallel to the $B_{\parallel}$ component, avoiding orbital coupling due to the Lorentz force.

\begin{figure}[!b]
\centerline{\includegraphics[width = 9cm]{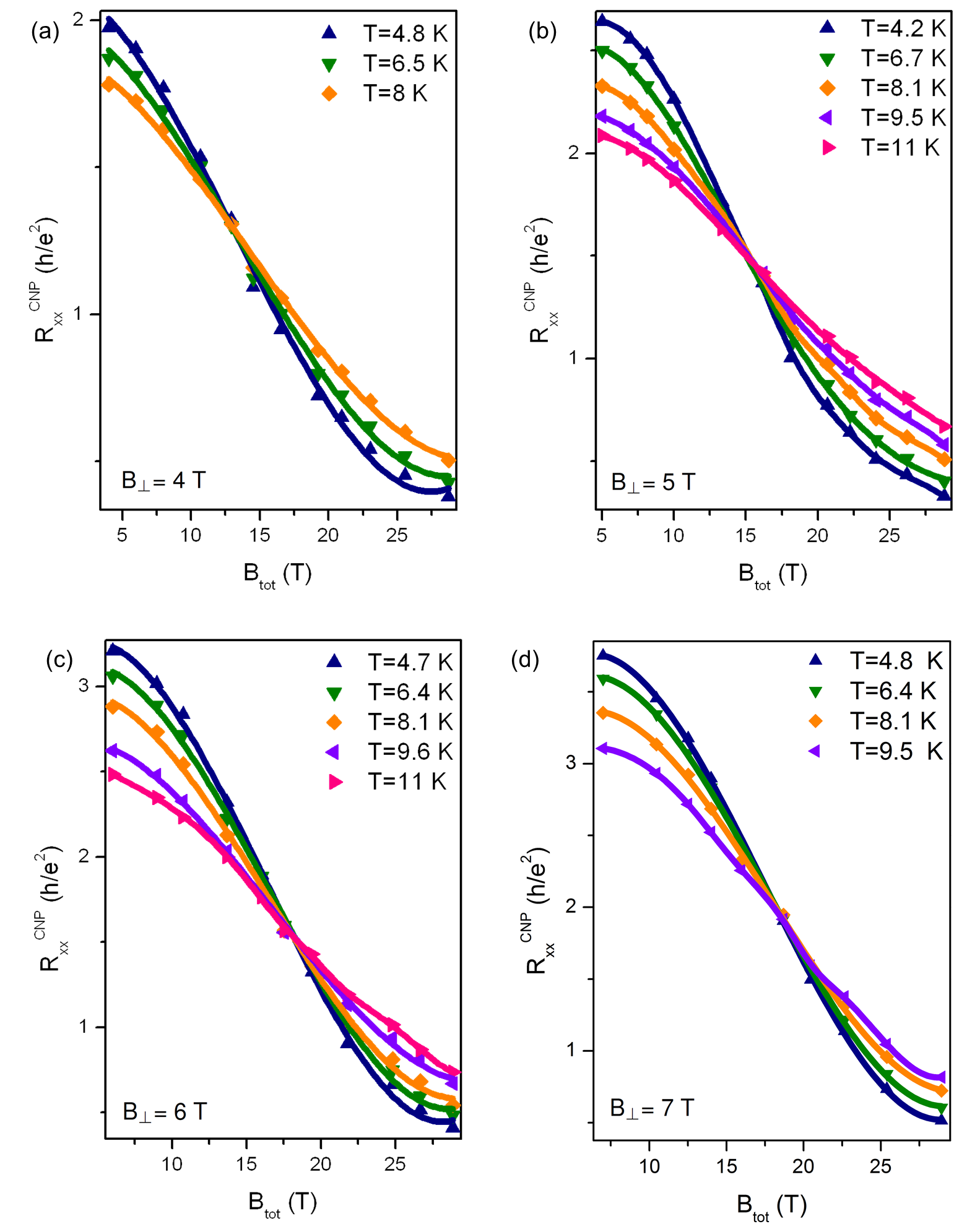}}
\caption{(Color online) Isotherms of $R_{xx}$ in correspondence of the CNP as a function of $B_{tot}$ with increasing perpendicular components: (a) $B_\perp=4$ T, (b) $B_\perp=5$ T, (c) $B_\perp=6$ T and (d) $B_\perp=7$ T. The symbols correspond to the experimental points, while the continuous lines are obtained by interpolation with grade-five polynomials.}
\label{two}
\end{figure}

At $B_{\parallel}=0$ T (\figurename~\ref{one} (a)) the CNP is individuated as an intense maximum close to $V_g=0$ V. It is worth commenting that higher values of $R^{CNP}_{xx}$ have been reported in other works with ultra-clean suspended samples \cite{yacoby}. In our supported sample some residual conductivity (possibly associated to charge puddles) is present even in this gapped phase, which is demonstrated by the clear thermally-activated behaviour shown in \figurename~\ref{one} (b) . Assuming that the resistance at the CNP follows $R_{xx}^{CNP}\propto \exp{(\Delta/2k_BT)}$,  we estimated an amplitude $\Delta=10.7\pm 0.4$ K for the energy gap at $B_\perp=7$ T, in good agreement with the results of Ref.\onlinecite{freitag}. Two metallic branches are symmetrically located with respect to the CNP, with values $R_{xx} <  h/e^2$ consistently with the population of the gapped edge channels associated to the CAF phase (analogously observed in Ref.\onlinecite{afy}). \figurename~\ref{one} (c) shows the evolution of $R_{xx}$ as a function of $V_g$ when a strong in-plane component ($B_\parallel$) is applied in addition to the quantizing perpendicular field $B_\perp$. As $B_\parallel$ increases $R^{CNP}_{xx}$ drops drastically, eventually evolving into a local minimum. The suppression of $R^{CNP}_{xx}$ is indicative of the creation of new conduction channels in the sample. When the highest values of $B_\parallel$ are considered, the local minimum at the CNP almost saturates and assumes values $\frac{1}{4}h/e^2 \lesssim R^{CNP}_{xx}<\frac{1}{2}h/e^2$, where the lower limit represents the expected value for the four-probe resistance in the F phase. The gap of the CAF phase actually closes at the edges of the sample, leading to four QSH-like helical edge states. The not exact quantization detected in our measurements (also observed in Ref.\onlinecite{maher,molen,konig}) indicates residual backscattering likely to be due to short-circuiting via bulk states and/or non-ideal contacts acting as uncontrolled de-coherence sites \cite{molen}. The metallic nature of the CNP in this configuration is clearly demonstrated by the $T$-dependent measurements shown in \figurename~\ref{one} (d), indicating the realization of an insulator-to-metal CAF-F phase transition.

\begin{figure}[!b]
\centerline{\includegraphics[width = 8.3cm]{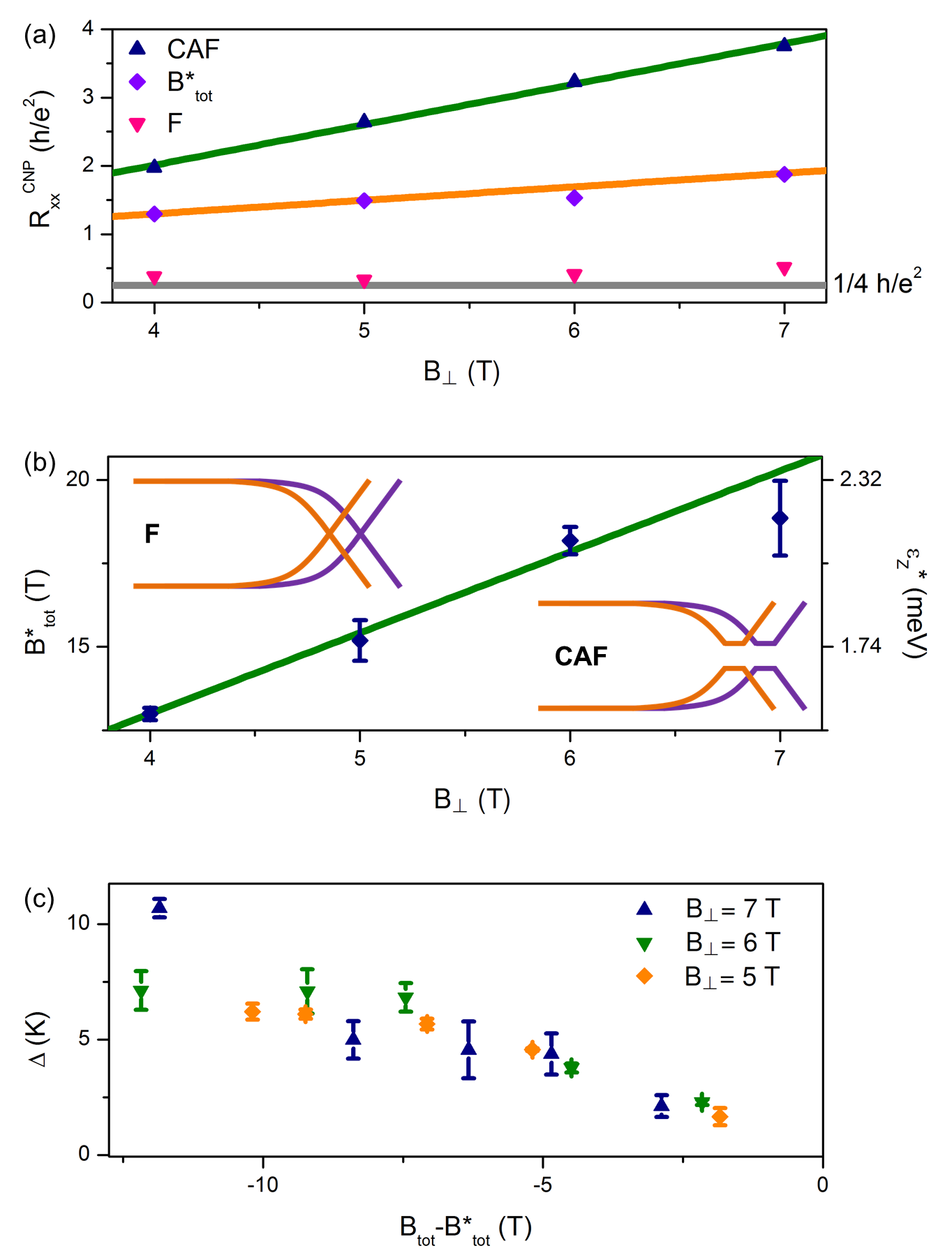}}
\caption{(Color online) (a)$R^{CNP}_{xx}$ as a function of $B_\perp$ in the absence of $B_\parallel$ (CAF, blue up triangles), with the largest $B_\parallel$ applied (F, pink down triangles) and at the critical point ($B^*_{tot}$, violet diamonds), at $T\approx 4.5$ K. The green line is a guide to the eye, the gray one indicates $\frac{1}{4}h/e^2$, while the orange one corresponds to $A(B_\perp)$ (see text). (b) Critical total field ($B^*_{tot}$, left axis) and corresponding critical Zeeman energy ($\epsilon^*_Z$, right axis) for the CAF-F transition as a function of $B_\perp$. The data points are calculated as mean values of $B_{tot}$ at the T-independent crossing points of the solid lines in Fig.\ref{two}, while the error bars correspond to the standard deviations. (inset) The bulk and edge energy spectrum of the CAF and F are represented on the two sides of the phase boundary. (c) Evolution of the energy gap $\Delta$ as a function of $B_{tot}-B^*_{tot}$, estimated from the thermally-activated behavior of $R^{CNP}_{xx}$ in the CAF phase at different values of $B_\perp$.}
\label{three}
\end{figure}

In order to individuate the critical field for the CAF-F transition demonstrated by the previous data, we measured $R^{CNP}_{xx}$ at different values of $T$ and $B_{tot}$, while keeping $B_\perp$ constant. This is done in order to keep the anisotropy energy $u_\perp$ unchanged, while increasing $\epsilon_Z$ and thus driving the transition. A $T$-independent crossing point for the isotherms of $R^{CNP}_{xx}$ is clearly visible in each panel of \figurename~\ref{two} (each of them showing data acquired at a single value $B_\perp=$ 4, 5, 6 and 7~T). These data provide the first experimental identification of the critical point for the CAF-F phase transition at CNP in bilayer graphene. In the insulating phase, the values of $R^{CNP}_{xx}$ clearly increase with $B_\perp$, in accordance with previous observations; \cite{yacoby,freitag} on the other hand $R^{CNP}_{xx}$ does not depend on $B_\perp$ in the metallic F phase (see \figurename~\ref{three} (a)). This indicates that the gap of the CAF phase increases with $B_\perp$, while the system becomes insensitive to $B_\perp$ once the F order is realized (the small deviations in the data at $B_\perp=6, 7$ T should become negligible if higher $B_{tot}$ could be reached with our experimental system).

The values of critical field $B^*_{tot}$ corresponding to the $T$-independent crossing points are plotted in \figurename~\ref{three} (b) as a function of $B_\perp$. It is clearly shown that $B^*_{tot}$ increases with $B_\perp$ and a linear dependence can be inferred. From a linear fit to the data (see green line in \figurename~\ref{three} (b)), we obtain
\begin{equation}
\label{eq2}
B^*_{tot}=(2.42\pm 0.21)\times B_\perp
\end{equation}
Equation \ref{eq2} represents the first experimental determination of the $B_\perp$-dependence for this critical field. If we consider the condition expressed by Equation \ref{trans}, we get an estimate for the anisotropy energy $u_\perp=(1.62\pm 0.14)$ K/T$\times B_\perp$ (assuming $g=2$), in accordance with the prediction of Ref.\onlinecite{kharprl,kharprb} ($u_\perp \approx 1- 10$ K/T$\times B_\perp$). The Zeeman energies at the critical points $\epsilon^*_Z$ (see right axis in \figurename~\ref{three} (b)) are consistent with the CAF-F boundary $\epsilon_Z\approx 0.8$ meV reported in Ref.\onlinecite{maher} for $B_\perp=1.75$ T, which has been estimated on the basis of the saturation of $R^{CNP}_{xx}$ rather than at the insulator-to-metal transition as done in the present work.
The reliability of our estimation of the critical point is further evidenced by the evolution of the energy gap $\Delta$ of the insulating phase as the in-plane field is increased. In \figurename~\ref{three} (c) we show that $\Delta$ (obtained by fitting the thermally-activated resistance as in \figurename~\ref{one} (b)) continuously decrease towards $B_{tot}=B^*_{tot}$, eventually pointing at vanishing values, perfectly matching the evolution of the edge gap described in Ref.\onlinecite{kharprb}. 

\begin{figure}[!ht]
\centerline{\includegraphics[width = 9cm]{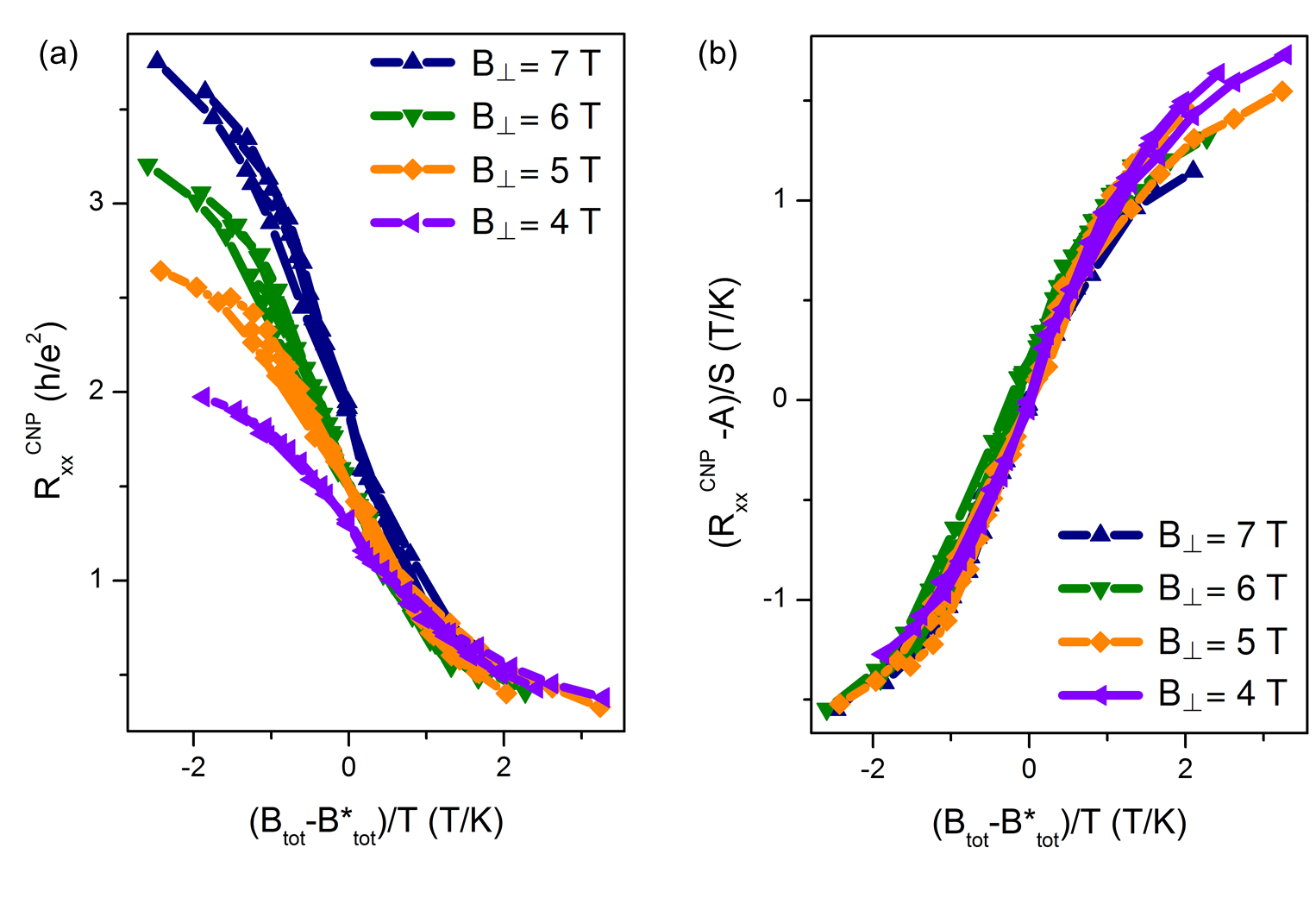}}
\caption{(Color online) (a) $R^{CNP}_{xx}$ as a function of $(B_{tot}-B^*_{tot})/T$ for the four values $B_\perp$ considered: scaling behavior is clearly shown. (b) $(R^{CNP}_{xx}-A(B_\perp))/S(B_\perp)$ as a function of $(B_{tot}-B^*_{tot})/T$, demonstrating the universality of the relation expressed by Equation \ref{eq3}.}
\label{four}
\end{figure}

A distinctive feature of the phase transitions in quantum systems is represented by the scaling of physical observables in the vicinity of the critical point at finite temperature \cite{sachdev}. In the following we individuate for the first time a scaling law for the CAF-F transitions observed at different $B_\perp$ and determine a universal expression for $R^{CNP}_{xx}$. \figurename~\ref{four} (a) shows the same data of $R^{CNP}_{xx}$ as in \figurename~\ref{two} (a), (b), (c) and (d), plotted as a function of $(B_{tot}-B^*_{tot})/T$. The data relative to the four values of $B_\perp$ considered collapse into four different curves, demonstrating a nearly ideal scaling of $R^{CNP}_{xx}$ with $(B_{tot}-B^*_{tot})/T$. The four curves coincide for $B_{tot}\gg B^*_{tot}$, where $R^{CNP}_{xx}$ has been shown to be a $B_\perp$-independent quantity, while they differ in the vicinity of the transition ($B_{tot}\approx B^*_{tot}$) and in the CAF region ($B_{tot}\ll B^*_{tot}$). For $B_{tot}\approx B^*_{tot}$, linearity is observed in the four curves. Our best estimate gives an intercept $A(B_\perp)=1.31\times 10^4+5.1\times 10 ^3B_\perp$(T) $\Omega$, together with a slope $S(B_\perp)=9.33\times 10^3+1.25\times 10 ^3B_\perp$(T) $\Omega$K/T, which apply for any of the $B_\perp$ values. These two parameters can therefore be inserted into a tentative general expression for the resistance at the CNP in the vicinity of the CAF-F transition:
\begin{equation}
\label{eq3}
R^{CNP}_{xx}(B_{tot},B_\perp,T)=A(B_\perp)+S(B_\perp)\left(\frac{B_{tot}-B^*_{tot}(B_\perp)}{T}\right)
\end{equation}
The universality of Equation \ref{eq3} is demonstrated in \figurename~\ref{four} (b), where $(R^{CNP}_{xx}-A(B_\perp))/S(B_\perp)$ is plotted as a function of $(B_{tot}-B^*_{tot})/T$ and our whole experimental data set collapses into a unique curve (as expected, Equation \ref{eq3} doesn't apply in the region $B_{tot}\gg B^*_{tot}$, where no dependence on $B_\perp$ has already been pointed out). The parameter $A(B_\perp)$ corresponds to the value of $R^{CNP}_{xx}$ at the critical field. The fact that $R^{CNP}_{xx}(B^*_{tot})$ is a $B_\perp$-dependent quantity is distinctive of this transition with respect to e.g. liquid-to-insulator transitions in two-dimensional electron gases \cite{sahar} (where a universal value of the resistance is found at the critical point), and once more points out the role of the anisotropy energy $u_\perp$ in this kind of phenomenology.

In conclusion, we performed magnetotransport measurements on a high-mobility bilayer graphene sample enclosed by $h$-BN, especially making use of tilted magnetic fields. We found clear evidence of a transition from an insulating to a metallic state at the CNP, which is driven by the application of a strong in-plane component of the magnetic field and is consistent with a CAF-F spin phase transition. We showed for the first time that a critical point for this transition can be individuated, corresponding to a $T$-independent crossing for the isotherms of the four-probe resistance and obtained a linear dependence for the critical field as a function of the perpendicular component. In the same way, the value of the four-probe resistance at the transition is found to depend on the perpendicular component of the field, which determines the strength of the interaction-induced anisotropy. In the vicinity of the transition we demonstrated scaling behavior for the resistance and we determined an universal expression as a function of total magnetic field, perpendicular component and temperature. The latter findings of our analysis are of particular relevance in encouraging further theoretical work on the subject. 

We acknowledge Dr. Niko Tombros for providing the sample and for enlightening discussions. This work has been supported by the following projects: JCYL SA226U13, FPU AP2009-2619 and European Union CTA-228043-EuroMagNET II Programme.

\end{document}